\documentclass[letterpaper]{ptephy_v1}
\preprintnumber{XXXX-XXXX} 

\usepackage{url}
\usepackage{dcolumn}

\begin{document}

\title{Observation~of~a~Be~double-Lambda hypernucleus in~the~J-PARC E07 experiment}


\author[1,2]{H.~Ekawa}
\author[3]{K.~Agari}
\author[4]{J.~K.~Ahn}
\author[5]{T.~Akaishi}
\author[3]{Y.~Akazawa}
\author[1,2]{S.~Ashikaga}
\author[6]{B.~Bassalleck}
\author[7]{S.~Bleser}
\author[8]{Y.~Endo}
\author[1]{Y.~Fujikawa}
% ...

\affil[1]{Department of Physics, Kyoto University, Kyoto 606-8502, Japan}
\affil[2]{Advanced Science Research Center, Japan Atomic Energy Agency, Tokai 319-1195, Japan}
\affil[3]{Institute of Particle and Nuclear Study (IPNS), High Energy Accelerator Research Organization (KEK), Tsukuba 305-0801, Japan}
\affil[4]{Department of Physics, Korea University, Seoul 02841, Korea}
\affil[5]{Department of Physics, Osaka University, Toyonaka 560-0043, Japan}
\affil[6]{Department of Physics and Astronomy, University of New Mexico, Albuquerque, New Mexico 87131, USA}
\affil[7]{Helmholtz Institute Mainz, 55099 Mainz, Germany}
\affil[8]{Physics Department, Gifu University, Gifu 501-1193, Japan}
% ...


\begin{abstract}
  A double-$\Lambda$ hypernucleus, ${}_{\Lambda\Lambda}\mathrm{Be}$, was observed by the J-PARC E07 collaboration
  in nuclear emulsions tagged by the $(K^{-},K^{+})$ reaction.
  This event was interpreted as a production and decay of $ {}_{\Lambda\Lambda}^{\;10}\mathrm{Be}$,
  ${}_{\Lambda\Lambda}^{\;11}\mathrm{Be}$, or ${}_{\Lambda\Lambda}^{\;12}\mathrm{Be}^{*}$ via $\Xi^{-}$ capture in ${}^{16}\mathrm{O}$.
  By assuming the capture in the atomic 3D state, the binding energy of two $\Lambda$ hyperons$\,$($B_{\Lambda\Lambda}$)
  of these double-$\Lambda$ hypernuclei are obtained to be
  $15.05 \pm 0.11\,\mathrm{MeV}$, $19.07 \pm 0.11\,\mathrm{MeV}$, and $13.68 \pm 0.11\,\mathrm{MeV}$, respectively.
  Based on the kinematic fitting, ${}_{\Lambda\Lambda}^{\;11}\mathrm{Be}$ is the most likely explanation for the observed event.
\end{abstract}

\subjectindex{D14}

\maketitle

\section{Introduction}
A complete understanding of the baryon-baryon interaction requires (at least) the consideration of all octet baryons within the SU(3)$_{f}$ group.
Compared to baryon pairs with zero or only one strange baryon, experimental information in the $S = -2$ sector is still very scarce.
In particular, it is difficult to study the interaction between two hyperons by scattering experiments due to their short lifetimes.
Therefore, double-$\Lambda$ hypernuclei, which include two $\Lambda$ hyperons in a nucleus, have been investigated.
The $\Lambda\Lambda$ interaction is expressed in terms of $\Delta B_{\Lambda\Lambda}$, which can be deduced from the mass of a double-$\Lambda$ hypernucleus
and which is defined as
\begin{equation}
  \Delta B_{\Lambda\Lambda}({}^{\;\;A}_{\Lambda\Lambda}Z) = B_{\Lambda\Lambda}({}^{\;\;A}_{\Lambda\Lambda}Z) - 2B_{\Lambda}({}^{A-1}_{\;\;\;\;\:\,\Lambda}Z).
\end{equation}
Here $B_{\Lambda}$ and $B_{\Lambda\Lambda}$ represent the binding energies of a $\Lambda$ hyperon in single-$\Lambda$ hypernuclei
and two $\Lambda$ hyperons in double-$\Lambda$ hypernuclei, respectively.
Emulsion detectors allow the detection of the weak decay products from double-$\Lambda$ hypernuclei with sub-$\mathrm{\mu m}$ resolution,
thus providing the most precise reconstruction of double-$\Lambda$ hypernucleus masses as of today.

In the past, several experiments have successfully discovered double-$\Lambda$ hypernuclear decays in nuclear emulsions~\cite{danysz, e176, takahashi, ahn}.
The most impressive results were collected by the KEK-PS E373 experiment.
Among seven double-$\Lambda$ hypernuclear events, an event called ``NAGARA'' was uniquely identified as ${}_{\Lambda\Lambda}^{\;\;\:6}\mathrm{He}$~\cite{takahashi}.
From this event, the $\Lambda\Lambda$ interaction, especially its $s$-wave (${}^{1}S_{0}$) interaction,
was found to be weakly attractive ($\Delta B_{\Lambda\Lambda} = 0.67 \pm 0.17\,\mathrm{MeV}$)~\cite{ahn}.
In order to understand $\Lambda\Lambda$ and the related interaction systematically, more double-$\Lambda$ hypernuclei need to be uniquely identified.

J-PARC E07 is an upgraded counter-emulsion hybrid experiment aiming to identify about a factor of
10 more double-$\Lambda$ hypernuclei as compared to the E373 experiment~\cite{e07}.
It is expected to detect approximately 100 double-$\Lambda$ hypernuclear events among $1 \times 10^{4}\;\Xi^{-}$ stopping events.
This substantially improved statistics will allow us to explore not only the $\Lambda\Lambda$ $s$-wave interaction
but also $\Xi N$-$\Lambda\Lambda$ mixing, and the structure of the core nucleus, etc.

Beam exposure of E07 was carried out in 2016 and 2017.
A total of 118 modules produced from 2.1 tons emulsion gel were exposed to $1.13 \times 10^{11}$ particles of the $K^{-}$ beam.
An impressive double-$\Lambda$ hypernuclear event, ``MINO'', was observed after scanning 30\% of all modules.
In this paper, an interpretation of this event is discussed.

\section{Experimental setup}
\begin{figure}
  \centering\includegraphics[width=3.2in]{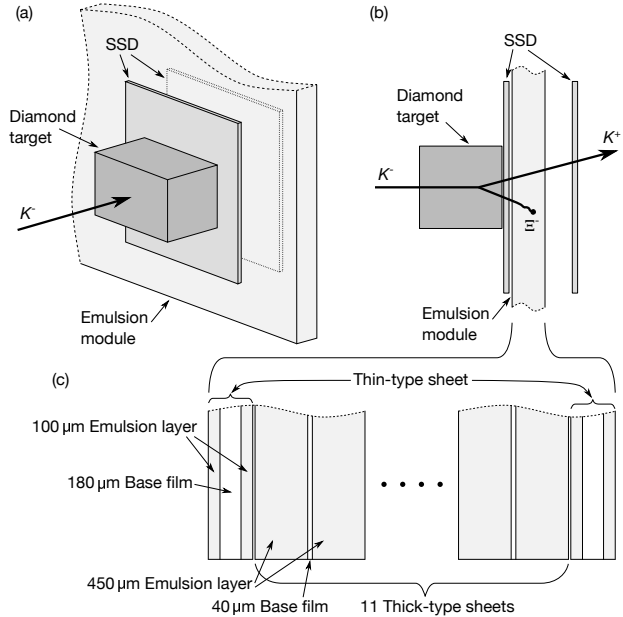}
  \caption{Schematic view of experimental setup around a diamond target.\
  The target size was $50\,\mathrm{mm}~[\mathrm{W}] \times 30\,\mathrm{mm}~[\mathrm{H}] \times 30\,\mathrm{mm}~[\mathrm{T}]$.
  }
  \label{emulsion}
\end{figure}

Double-$\Lambda$ hypernuclei were searched for by detecting $\Xi^{-}$ stopping events in an emulsion module.
A $\Xi^{-}$ hyperon generated in the quasi-free $(K^{-},K^{+})$ reaction in a diamond target
with $30\,\mathrm{mm}$ thickness~($9.83\,\mathrm{g/cm^{2}}$) was injected into the module and
subsequently slowed down and captured in the atomic orbit of a nucleus in the emulsion material.
Double-$\Lambda$ hypernuclei are produced by the interaction between the $\Xi^{-}$ hyperon and nucleus with a probability of a few percents.
In order to trace $\Xi^{-}$ tracks to the module and detect particle tracks from the module,
two sets of Silicon Strip Detector (SSD) were installed so as to sandwich the module.
Experimental setup around the target is shown in Fig.~\ref{emulsion}.

J-PARC E07 experiment was performed at K1.8 beam line in the J-PARC Hadron Experimental Facility with $K^{-}$ beams of $1.8\,\mathrm{GeV}/c$ momentum.
This momentum was chosen to maximize the $\Xi^{-}$ stopping yield in the emulsion.
Typical intensity and purity of the $K^{-}$ beam were $2.8 \times 10^{5}$ particles per spill of $2.0\,\mathrm{s}$ duration every $5.5\,\mathrm{s}$ and 82\%, respectively.
The momenta of incoming $K^{-}$ and outgoing $K^{+}$ mesons were analyzed by corresponding magnetic spectrometers,
the beam line spectrometer~\cite{k18} and the KURAMA spectrometer, respectively.
Momentum resolution of each spectrometer was $\Delta p/p = 3.3 \times 10^{-4}\mathrm{(FWHM)}$ and $\Delta p/p = 2.7 \times 10^{-2}\mathrm{(FWHM)}$, respectively.
The acceptance of the KURAMA spectrometer was $280\;\mathrm{msr}$.

Each emulsion module consisted of eleven thick-type sheets sandwiched between two thin-type sheets
with an area of $345\,\mathrm{mm}\,[\mathrm{W}] \times 350\,\mathrm{mm}\,[\mathrm{H}]$ [Fig. \ref{emulsion}(c)].
The thin-type sheets had emulsion layers with a thickness of $100\,\mathrm{\mu m}$ on both sides of $180\,\mathrm{\mu m}$ polystyrene base film and
were used to connect tracks to the SSD's, because they have high deformation tolerance thus good angular resolution.
The thick-type sheets had $450\,\mathrm{\mu m}$ thick layers on both sides of $40\,\mathrm{\mu m}$ polystyrene base film.
The emulsion layers were made of ``Fuji GIF'' emulsion gel produced by FUJIFILM Corporation.
The thirteen sheets were packed all together in a stainless case and fixed tightly by vacuum pumping.

The module was moved in the beam spill-off period to keep the beam particles density
less than $1 \times 10^{4}\,\mathrm{particles}/\mathrm{mm}^{2}$ in order to keep good efficiency for automated image tracking.
Typical exposure cycle was 5 hours for one module.

Each SSD had a four layers configuration (XYXY) with a strip pitch of $50\,\mathrm{\mu m}$.
The thickness of a silicon sensor was $0.3\,\mathrm{mm}$.
The resolutions of SSD's were estimated to be $15\,\mathrm{\mu m}\,(\sigma)$ for position and $20\,\mathrm{mrad}\,(\sigma)$ for angle.

\section{Analysis}
Tracks of $\Xi^-$ were kinematically identified with the SSD by tagging the $(K^{-},K^{+})$ reaction.
The candidate tracks of $\Xi^-$ hyperons were constructed from hits in four layers checking the consistency with the $\mbox{`}p\mbox{'}(K^{-},K^{+})\Xi^{-}$ kinematics.
In order to select $\Xi^-$ hyperons having high stopping probability, the $\Xi^-$ candidates which have more than six times larger energy deposit than
the minimum ionizing particles in the SSD were selected.
The downstream SSD was used to reject $\Xi^{-}$ candidates which penetrated the module without nuclear interactions.
A relative position between the module and the SSD's was calibrated using $\overline{p}$ beam through events with a beam-pattern matching method.
By calibrating four corners of the module, it was corrected with $20\,\mathrm{\mu m}$ accuracy.
Prediction accuracy of $\Xi^{-}$ incident position  was estimated to be about $50\,\mathrm{\mu m}$ for the thin-type sheet.

From the sets of the predicted positions and angles of the $\Xi^{-}$ hyperons based on the SSD hits,
the tracks were traced through the emulsion with automated microscope systems~\cite{soe}.
At first, the most upstream sheet of the module was scanned to find $\Xi^-$ tracks in an area of about $200\,\mathrm{\mu m} \times 200\,\mathrm{\mu m}$
for each $\Xi^{-}$ hit position predicted in the previous step of analysis.
Then, the $\Xi^{-}$ tracks found in the first thin-type sheet were traced downstream through several emulsion sheets.
When the tracing by the microscope system detected the end point of the track, the system took photographs around the stopping point
and those were checked by human eyes.

For kinematic analysis, range-energy calibration and shrinkage correction were necessary and performed for each emulsion sheet by
$\alpha$ tracks with the monochromatic energy of $8.784\,\mathrm{MeV}$
from the decay of ${}^{212}\mathrm{Po}$ existing in the emulsion.
The $\alpha$ track can be identified in the thorium series isotopes because it has the largest kinetic energy, ${\it i.e.,}$ the longest range.
Such $\alpha$ decay chains were searched for around the observed event by using the so called Overall scanning method~\cite{yoshida}.
One hundred $\alpha$ tracks were scanned for calibration.
Since the emulsion layers were shrunk along the beam direction due to photographic development,
the ranges of particles needed to be corrected for the shrinkage effect.
The mean range of $\alpha$ tracks and the shrinkage factor were obtained to be
$50.77 \pm 0.12\,\mathrm{\mu m}$ and $1.98 \pm 0.02$, respectively.
The relation between ranges and kinetic energies of charged particles was obtained by the range-energy formula given by Barkas {\it et. al.},~\cite{barkas,heckman}.
The density of the emulsion sheet was determined to be $3.486 \pm 0.013\,\mathrm{g/cm^{3}}$.
Ranges of particles in polystyrene base films and SSD's were converted into corresponding emulsion ranges considering the energy loss ratio.

\section{Interpretation of MINO event}
\begin{figure}
  \centering\includegraphics[width=3.5in]{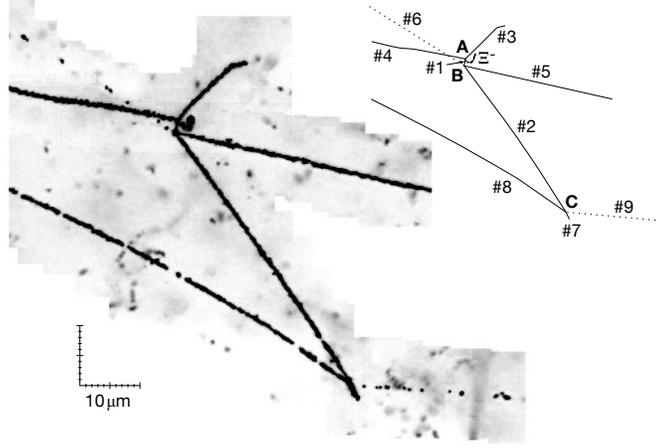}
  \caption{A photograph of the MINO event and its schematic drawing. The overlaid photograph is made by patching focused regions.\
  Tracks \#4, \#5, \#6, \#8, and \#9 are not fully shown in this photograph because these tracks are too long to be presented. }
  \label{event}
\end{figure}

A new double-$\Lambda$ hypernuclear event was observed in the 7th sheet of a module.
An overlaid photograph and a schematic drawing of the event are shown in Fig.~\ref{event}.
We named this event ``MINO''\footnote{The name of the southern part of Gifu prefecture, Japan, where the event was found.}.

The $\Xi^{-}$ hyperon came to rest at vertex A, from which three charged particles (\#1, \#3, and \#4) were emitted.
The particle of track \#1 decayed to three charged particles (\#2, \#5, and \#6) at vertex B.
The particle of track \#2 decayed again to three charged particles (\#7, \#8, and \#9) at vertex C.
Measured ranges and angles are summarized in Table~\ref{measurement}.
If the $\Xi^{-}$ hyperon was captured in a heavy nucleus such as Ag or Br,
a short track like \#3 with a range of less than $32\,\mathrm{\mu m}$ could not be emitted due to the Coulomb barrier~\cite{e176}.
Therefore, we have concluded that the $\Xi^{-}$ hyperon was captured in a light nucleus such as ${}^{12}\mathrm{C}$, ${}^{14}\mathrm{N}$, or ${}^{16}\mathrm{O}$.
The particles of tracks \#6 and \#9 escaped from the module into the downstream SSD after passing several emulsion sheets.
These tracks could be connected to the SSD by extrapolating the tracks at the exit point from the last emulsion sheet.
The particle of track \#6 was found to be stopped in the SSD 4th layer and \#9 penetrated all SSD layers.
The ranges of \#6 and \#9 in the SSD were $4500 \pm 200\,\mathrm{\mu m}$ and $2200 \pm 20\,\mathrm{\mu m}$ in emulsion equivalent, respectively.

\begin{table}
  \caption{Ranges and angles of the tracks of the MINO event. The zenith and azimuthal angles are presented in columns $\theta$ and $\phi$. }
  \label{measurement}
  \centering
  \begin{tabular}{c >{\centering}p{15mm} D{,}{\;\pm\;}{4} D{,}{\;\pm\;}{4} D{,}{\;\pm\;}{3} l}
    \hline
    \multicolumn{1}{>{\centering}p{9mm}}{Vertex}    &
    \multicolumn{1}{>{\centering}p{15mm}}{Track ID} &
    \multicolumn{1}{c}{Range$\,$[$\mathrm{\mu m}$]} &
    \multicolumn{1}{c}{$\theta$$\,$[degree]}        &
    \multicolumn{1}{c}{$\phi$$\,$[degree]}          &
    \multicolumn{1}{l}{Comment}                     \\
    \hline
    \hline
    \;\;A & \#1 & 2.1    , 0.2  & 83.7  , 8.9  & 256.1 , 5.3 & double-$\Lambda$ hypernucleus     \\
          & \#3 & 17.5   , 0.2  & 121.9 , 1.9  & 48.2  , 1.3 &                                   \\
          & \#4 & 65.7   , 0.5  & 41.7  , 1.7  & 166.7 , 2.1 &                                   \\
    \hline
    \;\;B & \#2 & 50.6   , 0.3  & 90.2  , 2.0  & 306.3 , 1.3 & single-$\Lambda$ hypernucleus     \\
          & \#5 & 122.1  , 0.2  & 61.4  , 1.8  & 347.0 , 1.5 &                                   \\
          & \#6 & \multicolumn{1}{c}{$>\!23170\quad$} & 106.2 , 0.6  & 147.7 , 0.4 & stopped in the SSD     \\
    \hline
    \;\;C & \#7 & 5.0    , 0.2  & 31.1  , 2.8  & 297.0 , 4.0 &                                   \\
          & \#8 & 116.7  , 0.2  & 100.3 , 1.9  & 144.2 , 1.3 &                                   \\
          & \#9 & \multicolumn{1}{c}{$>\!7378\quad$}  & 147.4 , 0.3  & 355.7 , 0.5 & passed through the SSD \\
    \hline
  \end{tabular}
\end{table}

We began with checking vertex C.
The coplanarity is defined as $(\overrightarrow{r_{1}} \times \overrightarrow{r_{2}}) \cdot \overrightarrow{r_{3}}$,
where $\overrightarrow{r_{i}}$ is a unit vector of a track angle.
Three charged particles were emitted with a coplanarity of $0.001 \pm 0.043$.
It shows that three particles were emitted in a plane; thus, neutron emission is unlikely.
The possibility of neutron emission is discussed in the end of this section.
From all nuclide combinations for both mesonic and non-mesonic decays of known single-$\Lambda$ hypernuclei,
possible decay modes were selected using the following criteria.
(1) An angular difference between \#9 and the momentum sum of \#7 and \#8 should be back-to-back with $3\,\sigma$ confidence.
(2) Momenta and energies should be conserved with $3\,\sigma$ by applying the kinematic fitting with degree of freedom (DOF) of 3~\cite{avery}.
Here, the range of \#9 was parameterized to conserve the total momentum and reconstruct the mass of a single-$\Lambda$ hypernucleus.
Possible decay modes at vertex C are listed in Table~\ref{vertexC}.
When the chi-square value of the kinematic fitting was larger than 14.2, such decay modes were rejected.
In this case, the p-value for the fitting is evaluated as 0.27\%, which corresponds to the $3\,\sigma$ cut condition.
Taking this into account, the possible candidate of \#2 was identified to be ${}_{\Lambda}^{5}\mathrm{He}$ in the case of no neutron emission.
The lower limit of the range of \#9 was obtained to be $7378\,\mathrm{(in\ emulsion)} + 2200\,\mathrm{(in\ SSD)}\,\mathrm{\mu m}$
considering the track length in the SSD.
The interpretation of ${}_{\Lambda}^{5}\mathrm{He}$ is consistent with this requirement.

\begin{table}
  \caption{Possible decay modes at vertex C in the case of no neutron emission.\
  Candidates which are accepted by the angular constraint and the conservation of momentum and energy\
  in $3\,\sigma$ cut condition are listed.\
  The $\chi^{2}$ value and the total range of \#9 were obtained from the kinematic fitting~\cite{avery}.}
  \label{vertexC}
  \centering
  \begin{tabular}{cccccccl}
    \hline
    Single-$\Lambda$ hypernucleus (\#2) & & \#7 & \#8 & \#9 & $\chi^{2}$ & Range$\,$(\#9)$\,$[$\mathrm{\mu m}$] & Comment \\
    \hline
    \hline
    ${}_{\Lambda}^{4}\mathrm{He}$ & $\to$ & ${}^{3}\mathrm{He}$ & $p$ & $\pi^{-}$ & 33.1 & 16800 & rejected \\
    ${}_{\Lambda}^{5}\mathrm{He}$ & $\to$ & ${}^{4}\mathrm{He}$ & $p$ & $\pi^{-}$ & 5.23 & 16270 &          \\
    ${}_{\Lambda}^{8}\mathrm{Li}$ & $\to$ & ${}^{6}\mathrm{Li}$ & $d$ & $\pi^{-}$ & 93.6 & 7906  & rejected \\
    ${}_{\Lambda}^{9}\mathrm{Li}$ & $\to$ & ${}^{7}\mathrm{Li}$ & $d$ & $\pi^{-}$ & 105  & 10660 & rejected \\
    \hline
  \end{tabular}
\end{table}

Next, we checked vertex B.
The particle of track \#1 decayed to three charged particles including a very thin track$\,$(\#6).
The range of \#6 was measured to be 23170 (in the emulsion) + 4500 (in the SSD)$\,\mathrm{\mu m}$.
If the particle of track \#6 is $\pi^{-}$, the total visible energy by decay daughters (\#2, \#5, and \#6) is at least $47.7\,\mathrm{MeV}$.
Since this energy is larger than the Q value of any $\pi^{-}$ mesonic decay mode, the possibility of \#6 to be $\pi^{-}$ was rejected.
Thus, the charge of \#1 should be more than three.
The maximum charge of \#1 is five by assuming $\Xi^{-}$ was captured in ${}^{16}\mathrm{O}$ because three charged particles were emitted at vertex A.
Therefore, the nuclide of \#1 is ${}_{\Lambda\Lambda}\mathrm{Be}$ or ${}_{\Lambda\Lambda}\mathrm{B}$.

\begin{figure}
  \centering\includegraphics[width=1.7in]{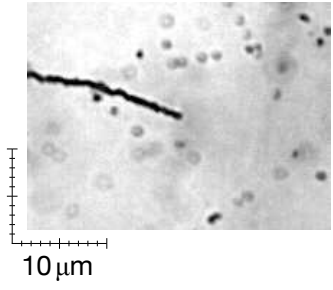}
  \caption{A photograph of the end point of track \#5.}
  \label{track5}
\end{figure}

The coplanarity of vertex B was calculated to be $0.007 \pm 0.019$.
Among the decay modes with ${}_{\Lambda}^{5}\mathrm{He}$ as \#2 without neutron emission,
only the following decay satisfied kinematical consistency.
\begin{equation*}
{}_{\Lambda\Lambda}^{\;13}\mathrm{B} \to {}_{\Lambda}^{5}\mathrm{He} + {}^{6}\mathrm{He} + d.
\end{equation*}
However, this decay mode was also rejected because there was no electron track associated with the end point of track \#5 as seen in Fig.~\ref{track5},
even though ${}^{6}\mathrm{He}$ should decay to ${}^{6}\mathrm{Li} + e^{-} + \overline{\nu}$ with a half-life of $806.7\,\mathrm{ms}$~\cite{tilley1}.
Thus, neutron(s) should be emitted at vertex B although the coplanarity is so small.
Regarding decay modes with neutron(s) at vertex B, all nuclide combinations were checked for charged particles.
In the kinematic analysis, the range of \#6 was calculated by assuming a double-$\Lambda$ hypernucleus with $\Delta B_{\Lambda\Lambda}=0$,
where the missing momentum was carried by unobserved neutron(s).
In the case of multiple neutron emissions, all neutrons were treated as having the same momentum.
This setting gives the minimum kinetic energy of neutrons and the maximum kinetic energy of \#6,
which corresponds to the maximum range of \#6.
If the maximum range was not consistent with the measurement, those assignments were rejected.
Since non-mesonic decays have large Q values, many decay modes remained for the case of \#1 being
${}_{\Lambda\Lambda}\mathrm{Be}$ or ${}_{\Lambda\Lambda}\mathrm{B}$ nuclides as summarized in Table~\ref{vertexB}.

\begin{table}
  \caption{Possible decay modes at vertex B.}
  \label{vertexB}
  \centering
  \begin{tabular}{cccccc}
    \hline
    Double-$\Lambda$ hypernucleus (\#1)   &       & \#2                           & \#5                                       & \#6 &          \\
    \hline
    \hline
    ${}_{\Lambda\Lambda}^{\;\;\:9}\mathrm{Be}$  & $\to$ & ${}_{\Lambda}^{5}\mathrm{He}$ & $p$                                       & $p$ & $2n$         \\
    ${}_{\Lambda\Lambda}^{\;10}\mathrm{Be}$     & $\to$ & ${}_{\Lambda}^{5}\mathrm{He}$ & $(p,d)$                                   & $p$ & $(3n,2n)$    \\
    ${}_{\Lambda\Lambda}^{\;11}\mathrm{Be}$     & $\to$ & ${}_{\Lambda}^{5}\mathrm{He}$ & $(p,d,t)$                                 & $p$ & $(4n,3n,2n)$ \\
    ${}_{\Lambda\Lambda}^{\;12}\mathrm{Be}$     & $\to$ & ${}_{\Lambda}^{5}\mathrm{He}$ & $(p,d,t)$                                 & $p$ & $(5n,4n,3n)$ \\
    ${}_{\Lambda\Lambda}^{\;13}\mathrm{Be}$     & $\to$ & ${}_{\Lambda}^{5}\mathrm{He}$ & $(p,d,t)$                                 & $p$ & $(6n,5n,4n)$ \\
    ${}_{\Lambda\Lambda}^{\;11}\mathrm{B}$      & $\to$ & ${}_{\Lambda}^{5}\mathrm{He}$ & ${}^{3}\mathrm{He}$                       & $p$ & $2n$         \\
    ${}_{\Lambda\Lambda}^{\;12}\mathrm{B}$      & $\to$ & ${}_{\Lambda}^{5}\mathrm{He}$ & $({}^{3}\mathrm{He}$,${}^{4}\mathrm{He})$ & $p$ & $(3n,2n)$    \\
    ${}_{\Lambda\Lambda}^{\;12}\mathrm{B}$      & $\to$ & ${}_{\Lambda}^{5}\mathrm{He}$ & ${}^{4}\mathrm{He}$                       & $d$ & $n$          \\
    ${}_{\Lambda\Lambda}^{\;13}\mathrm{B}$      & $\to$ & ${}_{\Lambda}^{5}\mathrm{He}$ & $({}^{3}\mathrm{He}$,${}^{4}\mathrm{He})$ & $p$ & $(4n,3n)$    \\
    ${}_{\Lambda\Lambda}^{\;14}\mathrm{B}$      & $\to$ & ${}_{\Lambda}^{5}\mathrm{He}$ & ${}^{4}\mathrm{He}$                       & $p$ & $4n$         \\
    ${}_{\Lambda\Lambda}^{\;15}\mathrm{B}$      & $\to$ & ${}_{\Lambda}^{5}\mathrm{He}$ & ${}^{4}\mathrm{He}$                       & $p$ & $5n$         \\
    \hline
  \end{tabular}
\end{table}

Finally, we checked vertex A, where three tracks were observed.
All nuclide combinations for \#1 to be ${}_{\Lambda\Lambda}\mathrm{Be}$ or ${}_{\Lambda\Lambda}\mathrm{B}$ were checked.
In case of the decay with neutron(s) emission, the momentum of neutron(s) was assumed to be the missing momentum.
In the case of more than one neutron emission, only a lower limit of $\Delta B_{\Lambda\Lambda}$ could be obtained.
Possible decay modes are listed in Table~\ref{vertexA}.
However, neutron(s) emission was unlikely because the coplanarity of vertex A was calculated to be $0.000 \pm 0.099$.
Additionally, $\Delta B_{\Lambda\Lambda} - B_{\Xi^{-}}$ should not be a large value considering the result of the NAGARA event.

\begin{table}
  \caption{Possible decay modes at Vertex A. Error of $\Delta B_{\Lambda\Lambda}$ shows the result of the kinematic fitting derived from our measurement.\
  Candidates which have $\Delta B_{\Lambda\Lambda} - B_{\Xi^{-}} < 20\,\mathrm{MeV}$ are listed.}
  \label{vertexA}
  \centering
  \begin{tabular}{ccccccD{,}{\;\pm\;}{7}}
    \hline
    \multicolumn{1}{c}{$\Xi^{-}$ capture} &
    \multicolumn{1}{c}{}                  &
    \multicolumn{1}{c}{\#1}               &
    \multicolumn{1}{c}{\#3}               &
    \multicolumn{1}{c}{\#4}               &
    \multicolumn{1}{c}{}                  &
    \multicolumn{1}{c}{$\Delta B_{\Lambda\Lambda} - B_{\Xi^{-}}\,\mathrm{[MeV]}$} \\
    \hline
    \hline
    ${}^{16}\mathrm{O} + \Xi^{-}$ & $\to$ & ${}_{\Lambda\Lambda}^{\;10}\mathrm{Be}$ & ${}^{4}\mathrm{He}$  & $t$ &      &    1.40   , 0.09   \\
    ${}^{16}\mathrm{O} + \Xi^{-}$ & $\to$ & ${}_{\Lambda\Lambda}^{\;11}\mathrm{Be}$ & ${}^{4}\mathrm{He}$  & $d$ &      &    1.64   , 0.08   \\
    ${}^{16}\mathrm{O} + \Xi^{-}$ & $\to$ & ${}_{\Lambda\Lambda}^{\;12}\mathrm{Be}$ & ${}^{4}\mathrm{He}$  & $p$ &      &    -2.95  , 0.08   \\
    ${}^{14}\mathrm{N} + \Xi^{-}$ & $\to$ & ${}_{\Lambda\Lambda}^{\;10}\mathrm{Be}$ & $p$                  & $p$ & $3n$ & >\!14.18  , 0.64   \\
    ${}^{14}\mathrm{N} + \Xi^{-}$ & $\to$ & ${}_{\Lambda\Lambda}^{\;10}\mathrm{Be}$ & $p$                  & $d$ & $2n$ & >\!15.94  , 1.22   \\
    ${}^{14}\mathrm{N} + \Xi^{-}$ & $\to$ & ${}_{\Lambda\Lambda}^{\;10}\mathrm{Be}$ & $d$                  & $p$ & $2n$ & >\!12.18  , 0.72   \\
    ${}^{14}\mathrm{N} + \Xi^{-}$ & $\to$ & ${}_{\Lambda\Lambda}^{\;10}\mathrm{Be}$ & $d$                  & $d$ &  $n$ &    16.29  , 1.92   \\
    ${}^{14}\mathrm{N} + \Xi^{-}$ & $\to$ & ${}_{\Lambda\Lambda}^{\;10}\mathrm{Be}$ & $t$                  & $p$ &  $n$ &    7.08   , 1.05   \\
    ${}^{14}\mathrm{N} + \Xi^{-}$ & $\to$ & ${}_{\Lambda\Lambda}^{\;11}\mathrm{Be}$ & $p$                  & $p$ & $2n$ & >\!10.60  , 0.93   \\
    ${}^{14}\mathrm{N} + \Xi^{-}$ & $\to$ & ${}_{\Lambda\Lambda}^{\;11}\mathrm{Be}$ & $p$                  & $d$ &  $n$ &    17.43  , 2.41   \\
    ${}^{14}\mathrm{N} + \Xi^{-}$ & $\to$ & ${}_{\Lambda\Lambda}^{\;11}\mathrm{Be}$ & $d$                  & $p$ &  $n$ &    10.30  , 1.38   \\
    ${}^{14}\mathrm{N} + \Xi^{-}$ & $\to$ & ${}_{\Lambda\Lambda}^{\;12}\mathrm{Be}$ & $p$                  & $p$ &  $n$ &    10.21  , 1.79   \\
    ${}^{16}\mathrm{O} + \Xi^{-}$ & $\to$ & ${}_{\Lambda\Lambda}^{\;10}\mathrm{Be}$ & ${}^{4}\mathrm{He}$  & $p$ & $2n$ & >\!9.35   , 0.47   \\
    ${}^{16}\mathrm{O} + \Xi^{-}$ & $\to$ & ${}_{\Lambda\Lambda}^{\;10}\mathrm{Be}$ & ${}^{4}\mathrm{He}$  & $d$ &  $n$ &    7.73   , 0.40   \\
    ${}^{16}\mathrm{O} + \Xi^{-}$ & $\to$ & ${}_{\Lambda\Lambda}^{\;11}\mathrm{Be}$ & ${}^{4}\mathrm{He}$  & $p$ &  $n$ &    4.42   , 0.91   \\
    ${}^{16}\mathrm{O} + \Xi^{-}$ & $\to$ & ${}_{\Lambda\Lambda}^{\;12}\mathrm{Be}$ & ${}^{3}\mathrm{He}$  & $p$ &  $n$ &    19.30  , 0.82   \\
    ${}^{16}\mathrm{O} + \Xi^{-}$ & $\to$ & ${}_{\Lambda\Lambda}^{\;13}\mathrm{B}$  & $p$                  & $p$ & $2n$ & >\!14.20  , 1.34   \\
    ${}^{16}\mathrm{O} + \Xi^{-}$ & $\to$ & ${}_{\Lambda\Lambda}^{\;13}\mathrm{B}$  & $d$                  & $p$ &  $n$ &    16.55  , 2.09   \\
    ${}^{16}\mathrm{O} + \Xi^{-}$ & $\to$ & ${}_{\Lambda\Lambda}^{\;14}\mathrm{B}$  & $p$                  & $p$ &  $n$ &    18.44  , 2.65   \\
    \hline
  \end{tabular}
\end{table}

From the above considerations, candidates for production and decay modes are following.
\begin{eqnarray*}
{}^{16}\mathrm{O} + \Xi^{-} \to  ({}_{\Lambda\Lambda}^{\;10}\mathrm{Be},\ {}_{\Lambda\Lambda}^{\;11}\mathrm{Be},\ {}_{\Lambda\Lambda}^{\;12}\mathrm{Be}) + {}^{4}\mathrm{He} + (t,\ d,\ p),\\
\hookrightarrow {}_{\Lambda}^{5}\mathrm{He} + (p,d,t) + p + xn,\quad\\
\hookrightarrow {}^{4}\mathrm{He} + p + \pi^{-}.\qquad\quad
\end{eqnarray*}
The nuclide of the double hypernucleus was uniquely identified as a ${}_{\Lambda\Lambda}\mathrm{Be}$.
The $B_{\Lambda\Lambda}$ and $\Delta B_{\Lambda\Lambda}$ values depend on the $\Xi^{-}$ binding energy ($B_{\Xi^{-}}$).
If we assume that the $\Xi^{-}$ hyperon was captured in the atomic 3D state of ${}^{16}\mathrm{O}$
with the theoretically estimated $B_{\Xi^{-}}$ value of 0.23 MeV~\cite{ehime},
$B_{\Lambda\Lambda}$ ($\Delta B_{\Lambda\Lambda}$) for each decay mode are obtained to be
$15.05 \pm 0.11\,\mathrm{MeV}\,(1.63 \pm 0.14\,\mathrm{MeV})$, $19.07 \pm 0.11\,\mathrm{MeV}\,(1.87 \pm 0.37\,\mathrm{MeV})$,
and $13.68 \pm 0.11\,\mathrm{MeV}\,(-2.7 \pm 1.0\,\mathrm{MeV})$, for
${}_{\Lambda\Lambda}^{\;10}\mathrm{Be}$, ${}_{\Lambda\Lambda}^{\;11}\mathrm{Be}$, and ${}_{\Lambda\Lambda}^{\;12}\mathrm{Be}$, respectively.
These values are summarized in Table~\ref{dBLL} together with their statistical and systematic errors.
The statistical error was caused by the kinematic fitting and the systematic error was caused by the mass of the $\Xi^{-}$ hyperon
and $B_{\Lambda}({}^{A-1}_{\;\;\;\;\:\,\Lambda}Z)$.
The mass of the $\Xi^{-}$ and $\Lambda$ hyperon was taken as $1321.71\pm0.07\,\mathrm{MeV}$ and $1115.683\pm0.006\,\mathrm{MeV}$, respectively~\cite{pdg}.
In this analysis, we took $8.2 \pm 0.5\,\mathrm{MeV}$ for ${B_{\Lambda}({}_{\;\,\Lambda}^{11}\mathrm{Be})}$
by a linear extrapolation from ${B_{\Lambda}}$ of ${}_{\Lambda}\mathrm{Be}$ isotopes
with values of $5.16 \pm 0.08\,\mathrm{MeV}$ (${}_{\Lambda}^{7}\mathrm{Be}$)~\cite{davis}, $6.84 \pm 0.05\,\mathrm{MeV}$ (${}_{\Lambda}^{8}\mathrm{Be}$)~\cite{davis},
$6.71 \pm 0.04\,\mathrm{MeV}$ (${}_{\Lambda}^{9}\mathrm{Be}$)~\cite{davis}, and $8.60 \pm 0.07 \pm 0.16\,\mathrm{MeV}$ (${}_{\;\,\Lambda}^{10}\mathrm{Be}$)~\cite{gogami}
because ${}_{\;\,\Lambda}^{11}\mathrm{Be}$ has not yet been observed.

\begin{table}
  \caption{ Result of $B_{\Lambda\Lambda}$ and $\Delta B_{\Lambda\Lambda}$ for MINO event.\
  The $\Xi^{-}$ hyperon was assumed to be captured in the atomic 3D state of ${}^{16}\mathrm{O}\,(B_{\Xi^{-}} = 0.23\,\mathrm{MeV})$.}
  \label{dBLL}
  \centering
  \begin{tabular}{cD{,}{\;\pm\;}{10}D{,}{\;\pm\;}{10}D{,}{\;\pm\;}{16}}
    \hline
    \multicolumn{1}{c}{Nuclide} &
    \multicolumn{1}{c}{$B_{\Lambda\Lambda}\,\mathrm{[MeV]}$} &
    \multicolumn{1}{c}{$\Delta B_{\Lambda\Lambda}\,\mathrm{[MeV]}$} &
    \multicolumn{1}{c}{$B_{\Lambda}({}^{A-1}_{\;\;\;\;\:\,\Lambda}Z)\,\mathrm{[MeV]}\quad\quad$} \\
    \hline
    \hline
    ${}_{\Lambda\Lambda}^{\;10}\mathrm{Be}$  & 15.05 , 0.09 \pm 0.07 &  1.63 , 0.09 \pm 0.11 & 6.71 , 0.04 \mbox{~\cite{davis}}          \\
    ${}_{\Lambda\Lambda}^{\;11}\mathrm{Be}$  & 19.07 , 0.08 \pm 0.07 &  1.87 , 0.08 \pm 0.36 & 8.60 , 0.07 \pm 0.16\mbox{~\cite{gogami}} \\
    ${}_{\Lambda\Lambda}^{\;12}\mathrm{Be}$  & 13.68 , 0.08 \pm 0.07 & (-2.7 , 0.08 \pm 1.0) & 8.2  , 0.5\;\mathrm{(extrapolation)}      \\
    \hline
  \end{tabular}
\end{table}

The probabilities of the three interpretations were evaluated with the chi-square value of the kinematic fitting with DOF of 3.
Chi-square and p-value for these three decay modes are summarized in Table~\ref{chi}.
It is found that the most probable interpretation of this event is ${}_{\Lambda\Lambda}^{\;11}\mathrm{Be}$ from the chi-square values.

\begin{table}
  \caption{ $\chi^{2}$ and p-value of the kinematic fitting at vertex A. }
  \label{chi}
  \centering
  \begin{tabular}{D{,}{\;\to\;}{16} c c}
    \hline
    \multicolumn{1}{c}{Decay mode}      &
    \multicolumn{1}{c}{$\chi^{2}$}      &
    \multicolumn{1}{c}{p-value$\,$[\%]} \\
    \hline
    \hline
    {}^{16}\mathrm{O} + \Xi^{-} , {}_{\Lambda\Lambda}^{\;10}\mathrm{Be} + {}^{4}\mathrm{He} + t  & 11.5 & 0.93 \\
    {}^{16}\mathrm{O} + \Xi^{-} , {}_{\Lambda\Lambda}^{\;11}\mathrm{Be} + {}^{4}\mathrm{He} + d  & 7.28 & 6.35 \\
    {}^{16}\mathrm{O} + \Xi^{-} , {}_{\Lambda\Lambda}^{\;12}\mathrm{Be} + {}^{4}\mathrm{He} + p  & 11.3 & 1.02 \\
    \hline
  \end{tabular}
\end{table}

In above analysis, it is assumed that no neutron was emitted at vertex C.
The interpretation of vertex C is important because the analysis of vertex~B is not effective
in selecting possible candidates from the kinematics due to the large Q values of non-mesonic decays.
If we assume neutron(s) emission, following decay modes are also accepted.
\renewcommand{\theequation}{\roman{equation}}
\setcounter{equation}{0}
\begin{align}
  {}_{\Lambda}^{3}\mathrm{H} &\to p + p +\pi^{-} + n, \label{eq1}\\
  {}_{\Lambda}^{4}\mathrm{H} &\to d + p +\pi^{-} + n. \label{eq2}
\end{align}
However, it should be noted that the branching ratio of ${}_{\Lambda}^{3}\mathrm{H}$ and ${}_{\Lambda}^{4}\mathrm{H}$ decay
measured in a past experiment show only less than 30 instances of decay mode (\ref{eq1}) and 5 of (\ref{eq2})
among about 2000 ${}_{\Lambda}^{3}\mathrm{H}$ and ${}_{\Lambda}^{4}\mathrm{H}$ decays, respectively~\cite{bertrand}.
Moreover, a theoretical calculation also supports this small decay probability, {\it e.g.,} 0.6\% for decay mode (\ref{eq1})~\cite{kamada}.
Thus, the ${}_{\Lambda\Lambda}\mathrm{Li}$ nuclides in vertex A is very unlikely.

\section{Discussion}
The newly observed double-$\Lambda$ hypernuclear event is interpreted as the production of a
${}_{\Lambda\Lambda}^{\;10}\mathrm{Be}$, ${}_{\Lambda\Lambda}^{\;11}\mathrm{Be}$, or ${}_{\Lambda\Lambda}^{\;12}\mathrm{Be}$ nucleus.
The present result was compared with the results of past experiments.
Candidates of ${}_{\Lambda\Lambda}\mathrm{Be}$ double-$\Lambda$ hypernuclei which were observed in past experiments are listed in Table~\ref{past}.

\begin{table}
  \caption{Summary of ${}_{\Lambda\Lambda}\mathrm{Be}$ double-$\Lambda$ hypernuclei observed in past experiments.\
  Multiple interpretations are listed in  MIKAGE and HIDA events.}
  \label{past}
  \centering
  \begin{tabular}{>{\centering}p{35mm} >{\centering}p{11mm} >{\centering}p{12mm} D{,}{\;\pm\;}{4} D{,}{\;\pm\;}{4} p{24mm}}
    \hline
     \multicolumn{1}{ >{\centering}p{35mm} }{Event}   &
     \multicolumn{1}{ >{\centering}p{11mm} }{Target}  &
     \multicolumn{1}{ >{\centering}p{12mm} }{Nuclide} &
     \multicolumn{1}{ c                    }{$B_{\Lambda\Lambda}\,\mathrm{[MeV]}$}          &
     \multicolumn{1}{ c                    }{\;$\Delta B_{\Lambda\Lambda}\,\mathrm{[MeV]}$} &
     \multicolumn{1}{ p{24mm}              }{Comment} \\
    \hline
    \hline
    MIKAGE~\cite{ahn}                & ${}^{12}\mathrm{C}$ & ${}_{\Lambda\Lambda}^{\;\;\:6}\mathrm{He}$   & 10.01 , 1.71 & 3.77 , 1.71 & \\
                                     & ${}^{12}\mathrm{C}$ & ${}_{\Lambda\Lambda}^{\;11}\mathrm{Be}$      & 22.15 , 2.94 & 3.95 , 3.00 & \\
                                     & ${}^{14}\mathrm{N}$ & ${}_{\Lambda\Lambda}^{\;11}\mathrm{Be}$      & 23.05 , 2.59 & 4.85 , 2.63 & \\
    \hline
    DEMACHIYANAGI~\cite{ahn}         & ${}^{12}\mathrm{C}$ & ${}_{\Lambda\Lambda}^{\;10}\mathrm{Be}^{*}$  & 11.90 , 0.13 & -1.52, 0.15 & most probable\\
    \hline
    HIDA~\cite{ahn}                  & ${}^{14}\mathrm{N}$ & ${}_{\Lambda\Lambda}^{\;12}\mathrm{Be}$      & 22.48 , 1.21 &             & \\
                                     & ${}^{16}\mathrm{O}$ & ${}_{\Lambda\Lambda}^{\;11}\mathrm{Be}$      & 20.83 , 1.27 & 2.61 , 1.34 & \\
    \hline
    Danysz~\cite{dalitz,davis2,annu} & ${}^{12}\mathrm{C}$ & ${}_{\Lambda\Lambda}^{\;10}\mathrm{Be}$      & 14.7  , 0.4  & 1.3  , 0.4  & \\
    \hline
  \end{tabular}
\end{table}

DEMACHIYANAGI event was interpreted as ${}_{\Lambda\Lambda}^{\;10}\mathrm{Be}^{*}$ (most probable), an excited state of ${}_{\Lambda\Lambda}^{\;10}\mathrm{Be}$,
with a $B_{\Lambda\Lambda}$ value of $11.90 \pm 0.13\,\mathrm{MeV}$~\cite{ahn}.
The present result of ${}_{\Lambda\Lambda}^{\;10}\mathrm{Be}$ interpretation is consistent with DEMACHIYANAGI event
when considering ${}_{\Lambda\Lambda}^{\;10}\mathrm{Be}$ was generated in the ground state.
The energy difference of the excited ($2_{1}^{+}$) and ground ($0_{1}^{+}$) states of the core nucleus
${}^{8}\mathrm{Be}$ is $3.03\,\mathrm{MeV}$~\cite{tilley2}, although both states are particle unbound.
The double-$\Lambda$ hypernucleus ${}_{\Lambda\Lambda}^{\;10}\mathrm{Be}$ with $B_{\Lambda\Lambda} = 14.7 \pm 0.4\,\mathrm{MeV}$  observed by Danysz
is also consistent with the present result~\cite{dalitz,davis2,annu}.

The present result of ${}_{\Lambda\Lambda}^{\;11}\mathrm{Be}$ interpretation is also consistent with the past results given by HIDA and MIKAGE,
in which $B_{\Lambda\Lambda}$ values were reported to be $22.15 \pm 2.94\,\mathrm{MeV}$, $23.05 \pm 2.59\,\mathrm{MeV}$, and $20.83 \pm 1.27\,\mathrm{MeV}$~\cite{ahn}.
The present result has a small error on $B_{\Lambda\Lambda}$ because no neutron was emitted at the production vertex of the double-$\Lambda$ hypernucleus.

In the case of ${}_{\Lambda\Lambda}^{\;12}\mathrm{Be}$, the present analysis shows negative $\Delta B_{\Lambda\Lambda}$ value ($-2.7 \pm 1.0\,\mathrm{MeV}$ ),
which is not consistent with NAGARA.
However, if ${}_{\Lambda\Lambda}^{\;12}\mathrm{Be}$ was produced in an excited state,
$\Delta B_{\Lambda\Lambda}$ of the ground state is increased by its excitation energy.
Therefore, the interpretation of ${}_{\Lambda\Lambda}^{\;12}\mathrm{Be}$ can not be excluded.
The core nucleus, ${}^{10}\mathrm{Be}$, has excited states 3.368--$6.263\,\mathrm{MeV}\,$~\cite{tilley2}.
The level of ${}_{\Lambda\Lambda}^{\;12}\mathrm{Be}$ is estimated to be similar to that of ${}^{10}\mathrm{Be}$~\cite{private1}.
Thus, $\Delta B_{\Lambda\Lambda}$ for the ground state of ${}_{\Lambda\Lambda}^{\;12}\mathrm{Be}$ becomes positive.

Three interpretations, ${}_{\Lambda\Lambda}^{\;10}\mathrm{Be}$, ${}_{\Lambda\Lambda}^{\;11}\mathrm{Be}$, and ${}_{\Lambda\Lambda}^{\;12}\mathrm{Be}^{*}$,
are accepted from the above consideration.
Among them, ${}_{\Lambda\Lambda}^{\;11}\mathrm{Be}$ is the most probable one from the analysis described in the previous section.

\section{Summary}
An experiment to search for double-$\Lambda$ hypernuclei with a counter-emulsion hybrid method, E07, was carried out at J-PARC.
An impressive double-$\Lambda$ hypernuclear event called ``MINO'' has been observed.
Based on kinematic analysis, the nuclide of the double-$\Lambda$ hypernucleus was uniquely identified as a ${}_{\Lambda\Lambda}\mathrm{Be}$.
The event was interpreted as one of the three following candidates.
\begin{align*}
  {}^{16}\mathrm{O} + \Xi^{-} &\to  {}_{\Lambda\Lambda}^{\;10}\mathrm{Be} + {}^{4}\mathrm{He} + t, \\
  {}^{16}\mathrm{O} + \Xi^{-} &\to  {}_{\Lambda\Lambda}^{\;11}\mathrm{Be} + {}^{4}\mathrm{He} + d, \\
  {}^{16}\mathrm{O} + \Xi^{-} &\to  {}_{\Lambda\Lambda}^{\;12}\mathrm{Be}^{*} + {}^{4}\mathrm{He} + p.
\end{align*}
$B_{\Lambda\Lambda}\,(\Delta B_{\Lambda\Lambda})$ of these double-$\Lambda$ hypernuclei was obtained to be
$15.05 \pm 0.11\,\mathrm{MeV}\,(1.63 \pm 0.14\,\mathrm{MeV})$, $19.07 \pm 0.11\,\mathrm{MeV}\,(1.87 \pm 0.37\,\mathrm{MeV})$,
and $13.68 \pm 0.11\,\mathrm{MeV}\,(-2.7 \pm 1.0\,\mathrm{MeV})$, respectively
by assuming the $\Xi^{-}$ capture in the atomic 3D state with $B_{\Xi^{-}}$ of $0.23\,\mathrm{MeV}$.
Negative $\Delta B_{\Lambda\Lambda}$ value of ${}_{\Lambda\Lambda}^{\;12}\mathrm{Be}$ indicates
it was produced in the excited state.
The most probable interpretation was found to be the production and decay of
the ${}_{\Lambda\Lambda}^{\;11}\mathrm{Be}$ nucleus from the kinematic fitting.

The emulsion scanning of the E07 experiment is ongoing.
Twice the statistics for $\Xi^{-}$ stopping events than that of E373 has been scanned and
more than ten events of double- and twin-$\Lambda$ hypernuclei have been observed up to the present.
Further impressive events are expected to be observed in the near future.

\ack
We thank the staff members of the J-PARC accelerator and the hadron experimental facility for providing beams with both excellent and stable conditions
and enough time in spite of trouble on the beam extraction device.
We are grateful to the staffs of FUJIFILM Corporation for producing our emulsion gel.
We thank KEK computing research center for providing computational resources.
We also thank the support of NII for SINET5.
We would like to express thanks to the staffs of the Kamioka Observatory, ICRR, for protecting our emulsion sheets in Kamioka mine against cosmic rays.
This experiment was supported by JSPS KAKENHI Grant Numbers 23224006 and 16H02180, MEXT KAKENHI Grant Numbers 15001001 (Priority Area)
and 24105002 (Innovative Area 2404), and NRF of Korea with Grant Number 2018R1A2B2007757. S.B., J.P. and F.S. are supported by DAAD PPP Japan 2017 57345296.



\begin{thebibliography}{99}

\bibitem{danysz}    M. Danysz {\it et al.}, Nuclear Physics {\bf 49}, 121 (1963).
\bibitem{e176}      S. Aoki {\it et al.}, Nucl. Phys. A {\bf 828}, 191 (2009).
\bibitem{takahashi} H. Takahashi {\it et al.}, Phys. Rev. Lett. {\bf 87}, 212502 (2001).
\bibitem{ahn}       J. K. Ahn {\it et al.}, Phys. Rev. C {\bf 88}, 014003 (2013).
\bibitem{e07}       K. Imai {\it et al.}, J-PARC E07 experiment. Systematic Study of Double-Strangeness System with an Emulsion-Counter Hybrid Method (Available at: \url{http://j-parc.jp/researcher/Hadron/en/pac_0606/pdf/p07-Nakazawa.pdf}, date last accessed November 15, 2018).
\bibitem{k18}       T. Takahashi {\it et al.}, Prog. Theor. Exp. Phys. {\bf 2012}, 02B010 (2012).
\bibitem{soe}       M. K. Soe, R. Goto, A. Mishina, Y. Nakanisi, D. Nakashima, J. Yoshida, and K. Nakazaw, Nucl. Instrum. Methods Phys. Res. Sect. A {\bf 848}, 66 (2017).
\bibitem{yoshida}   J. Yoshida, S. Kinbara, A. Mishina, K. Nakazawa, M. K. Soe, A. M. M. Theint, K. T. Tint, Nucl. Instrum. Methods Phys. Res. Sect. A {\bf 847}, 86 (2017).
\bibitem{barkas}    W. H. Barkas, Pure \& Applied Physics series, {\bf 15} I II (1963).
\bibitem{heckman}   H. H. Heckman, B. L. Perkins, W. G. Simon, F. M. Smith, and W. H. Barkas, Phys. Rev. {\bf 117}, 544, (1960).
\bibitem{avery}     Fitting Theory Writeups and References (Available at: \url{https://www.phys.ufl.edu/~avery/fitting.html}, date last accessed November 15, 2018).
\bibitem{tilley1}   D. R. Tilley, C. M. Cheves, J. L. Godwin, G. M. Hale, H. M. Hofmann, J. H. Kelley, C. G. Sheu, and H. R. Weller,  Nucl. Phys. A{\bf 708}, 3 (2002).
\bibitem{ehime}     M. Yamaguchi, K. Tominaga, Y. Yamamoto, and T. Ueda, Prog. Theor. Phys. {\bf 105}, 627 (2001).
\bibitem{pdg}       M. Tanabashi et al. (Particle Data Group), Phys. Rev. D {\bf 98}, 030001 (2018).
\bibitem{davis}     D. H. Davis, Contemporary Physics, {\bf 27:2}, 91 (1986).
\bibitem{gogami}    T. Gogami {\it et al.}, Phys. Rev. C {\bf 93}, 034314 (2016).
\bibitem{bertrand}  D. Bertrand, G. Coremans, C. Mayeur, J. Sacton, P. Vilain, G. Wilquet, J. H. Wickens, D. O' Sullivan, D. H. Davis, and J. E. Allen, Nucl.Phys. B {\bf 16} 77 (1970).
\bibitem{kamada}    H. Kamada, J. Golak, K. Miyagawa, H. Wita\l a, and W. Gl\"{o}ckle, Phys. Rev. C {\bf 57}, 1595 (1998).
\bibitem{tilley2}   D. R. Tilley, J. H. Kelley, J. L. Godwin, D. J. Millener, J. Purcell, C. G. Sheu, and H. R. Weller,  Nucl. Phys. A{\bf 745}, 155 (2004).
\bibitem{dalitz}    R. H. Dalitz, D. H. Davis, P. H. Fowler, A. Montwill, J. Pniewski, and J. A. Zakrzewski, Proc. R. Soc. A {\bf 426}, 1 (1989).
\bibitem{davis2}    D. H. Davis, Nucl. Phys. A {\bf 754}, 3c (2005).
\bibitem{annu}      E. Hiyama and K. Nakazawa, Annu. Rev. Nucl. Part. Sci. {\bf 68}, 131–59 (2018).
\bibitem{private1}  E. Hiyama, private communication.

\end{thebibliography}
%

\end{document}